\documentclass{article}

\usepackage{arxiv}

\usepackage[utf8]{inputenc} 
\usepackage[T1]{fontenc}    
\usepackage{hyperref}       
\usepackage{url}            
\usepackage{booktabs}       
\usepackage{amsfonts}       
\usepackage{nicefrac}       
\usepackage{microtype}      
\usepackage{lipsum}
\usepackage{multirow}
\usepackage{makecell}
\usepackage{graphicx}
\usepackage{amsmath}
\usepackage{subfigure}

\usepackage{mathrsfs}
\usepackage{colortbl}
\usepackage{xcolor}
\usepackage[authoryear]{natbib}

\usepackage{algorithm}
\usepackage{algpseudocode}
\usepackage{booktabs} 
\graphicspath{ {./images/} }

\newcommand{\ttt}[1]{\text{\texttt{#1}}}
\makeatletter
\renewcommand{\paragraph}{%
  \@startsection{paragraph}{4}%
  {\z@}{.75ex \@plus 1ex \@minus .2ex}{-1em}%
  {\normalfont\normalsize\bfseries}%
}
\makeatother

\title{Reinforcement Learning for Automated Cybersecurity Penetration Testing}

\author{
   \href{https://orcid.org/0009-0004-3565-3615}{\includegraphics[scale=0.06]{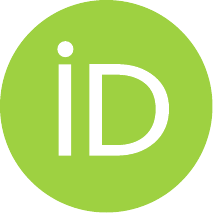}\hspace{1mm}}Daniel ~López-Montero \\
  Department of Artificial Intelligence and Big Data\\
  GMV\\
  Isaac Newton, 11, Tres Cantos, Madrid, Spain \\
  \texttt{daniel.lopez.montero@gmv.com} \\
   \And
 José L. ~Álvarez-Aldana \\
  Department of Artificial Intelligence and Big Data\\
  GMV\\
  Isaac Newton, 11, Tres Cantos, Madrid, Spain \\
  \texttt{jlalvarez@gmv.com} \\
  \And
   Alicia ~Morales-Martínez \\
  Department of Artificial Intelligence and Big Data\\
  GMV\\
  Isaac Newton, 11, Tres Cantos, Madrid, Spain \\
  \texttt{amorales.m@gmv.com} \\
  \And
 Marta ~Gil-López \\
  Department of Artificial Intelligence and Big Data\\
  GMV\\
  Isaac Newton, 11, Tres Cantos, Madrid, Spain \\
  \texttt{mglopez@gmv.com} \\
  \And
   \href{https://orcid.org/0000-0002-6158-4287}{\includegraphics[scale=0.06]{orcid.pdf}\hspace{1mm}}Juan M. ~Auñón-García \\
  Department of Artificial Intelligence and Big Data\\
  GMV\\
  Isaac Newton, 11, Tres Cantos, Madrid, Spain \\
  \texttt{jmaunon@gmv.com} \\
}

\begin{document}


\maketitle
\begin{abstract}
This paper aims to provide an innovative machine learning-based solution to automate security testing tasks for web applications, ensuring the correct functioning of all components while reducing project maintenance costs.
Reinforcement Learning is proposed to select and prioritize tools and optimize the testing path. The presented approach utilizes a simulated webpage along with its network topology to train the agent. Additionally, the model leverages Geometric Deep Learning to create priors that reduce the search space and improve learning convergence. The validation and testing process was conducted on real-world vulnerable web pages commonly used by human hackers for learning.
As a result of this study, a reinforcement learning algorithm was developed that maximizes the number of vulnerabilities found while minimizing the number of steps required.
\end{abstract}


\section{Introduction}
The rapid advancement of digital technology has significantly increased the complexity and scale of cyber threats. While traditional cybersecurity measures are effective to some extent, they often struggle to keep pace with the evolving tactics of sophisticated attackers \citep{saber_automated_2023}. In this landscape, penetration testing—wherein cybersecurity experts simulate attacks to identify vulnerabilities—has emerged as a critical component of any robust security strategy. However, manual penetration testing is time-consuming, resource-intensive, and often limited by the tester's knowledge and experience \citep{spieker_reinforcement_2017}.

Given these challenges, there is a growing emphasis on enhancing penetration testing (pentesting) through automation and intelligent decision-making. Pentesting plays a crucial role in assessing the security and resilience of a system by systematically identifying and exploiting vulnerabilities. This process involves gathering intelligence on potential targets, analyzing security weaknesses, and executing attacks to assess the system’s defenses. However, selecting the most effective course of action requires a high level of expertise, making the process both complex and time-consuming. While brute-force approaches could be employed—systematically testing all possible attack vectors—such methods are inefficient and can increase the risk of unintended consequences, such as Denial-of-Service (DoS) attacks that render network resources unavailable. To address these limitations, Artificial Intelligence (AI) is increasingly being explored as a means to enhance cybersecurity operations, offering solutions that can achieve performance comparable to that of human experts \citep{adawadkar_cyber-security_2022}.
\\
\\
One promising AI-driven approach is the application of Reinforcement Learning (RL) to pentesting, where an agent is trained to select the next action based on available information and network configurations. This decision-making process is guided by knowledge of the system under test, the outcomes of previous actions, and observations made during execution \citep{nguyen_pengym_2025}. By designing a reward function that prioritizes effective attack strategies, RL can help identify the most promising actions for compromising the system and achieving pentesting objectives.
The increasing integration of AI techniques, such as RL, reflects a broader trend in cybersecurity: the shift toward automation to handle complex security challenges. Over the past few decades, the field has experienced unprecedented growth, driven by advancements in technology, hardware, and algorithms, including machine learning. 
\\
\\
Traditionally, cybersecurity relied heavily on the expertise and direct intervention of specialized professionals. However, as threats have evolved and become more sophisticated, significant research efforts have focused on automating and streamlining tasks that were once performed exclusively by human experts. These efforts have resulted in several notable applications \citep{lopez-montero_towards_2023}:
(1) Automatically identifying the most promising attack paths based on network configuration and available information to discover new vulnerabilities.
(2) Estimating vulnerability risk by scoring a network's robustness while considering factors such as its graph structure (e.g., weakly or strongly connected) and initial configuration.
(3) Utilizing AI assistants to recommend actions for operators during penetration tests, helping them uncover potential vulnerabilities based on the current network state.
(4) Identifying the most critical vulnerabilities by balancing the impact of exploitation against the cost of remediation.
(5) Assessing cybersecurity risks and postures during mergers, acquisitions, or investments, which is crucial for due diligence processes. This includes identifying configurations that could enable attacks before deployment and measuring vulnerabilities arising from changes in network structure. Additionally, persistent techniques are used to combine opportunities—such as cached credentials and misconfigurations—into new attack paths.
\\
\\
\\
Most cybersecurity publications in the field of Machine Learning focus on anomaly and security breach detection \citep{shaukat_survey_2020}. However, the automation of pentesting tasks has also attracted a lot of interest recently \citep{almubairik_automated_2016, gangupantulu_using_2021}, the most outstanding being those that base their solution on Partial Observable Markov Decision Process (POMDP) \citep{schwartz_pomdp_2020} and Reinforcement Learning \citep{lopez-montero_towards_2023, hu_automated_2020, elderman_adversarial_2017}, approaches, as described below. The great contribution of this approach is the absence of human dependence with the main objective of automating the entire process, deciding at each moment the next action (scan, exploit, repeat...) in the design of an attack for vulnerability detection. The Reinforcement Learning (RL) technique is ideally suited for action prioritization tasks, which perfectly fits in the context of test automation. This prioritization can be included in Continuous Integration (CI) processes or failure detection during the execution of functional tests \citep{spieker_reinforcement_2017}. However, one of the use cases whose characteristics are best suited to RL is pentesting tasks \citep{ghanem_reinforcement_2020}. The main reason is that RL enables the execution of broader tests covering a wide range of attack vectors while also considering complex and evasive attack paths that are difficult for human testers to identify and analyze. Additionally, a network/topology simulation environment is used for training for two reasons: first, to prevent denial-of-service attacks on a real network during the process; and second, to ensure that the trained model has sufficient diversity to generalize effectively.
\\
\\
This manuscript is structured starting with a brief review of Penetration Testing and Reinforcement Learning; Section \ref{sec:environment} describes both, simulated and real environments, for training, validation, and testing. It includes a description of the action space, state, and reward feedback loop; Section \ref{sec:agent} shows the RL agent's technical details, as well as its connection to Geometrical Deep Learning and Graph Neural Networks; Section \ref{sec:experiments} describes the collection of experiments and results obtained; and finally, some conclusions and future work are presented in Section \ref{sec:results}.

\subsection{Penetration Testing}

Penetration testing, often referred to as a pentest, is a crucial method for safeguarding against cyber-crime. It involves simulating the actions of malicious hackers to identify and exploit potential vulnerabilities within a system. While pen-testing is sometimes confused with vulnerability assessment, they are distinct processes; vulnerability assessment focuses on identifying weaknesses, whereas penetration testing also involves exploiting them to gauge the extent of potential damage. The pen-test process is structured through a life cycle that begins with initial engagement and concludes with the delivery of a comprehensive final report, streamlining the testing process for greater efficiency \citep{saber_automated_2023}.

The penetration testing life cycle can be divided into the following consecutive different steps \citep{patil_ai-based_2024}:
 (1) Engagement:  A professional pentester defines the scope and fee for the test, considering factors like the type of engagement (White Box, Black Box, Grey Box), complexity, time requirements, and the number of targets (IP addresses, domains, etc.).
 (2) Information Gathering: This critical step involves collecting data about the target using sources like social networks, public websites, and tools like \textit{CrunchBase}, \textit{SAM}, and the \textit{Whois database}.
 (3) Footprinting and Scanning: With the gathered information, the pen-tester identifies active hosts and services within the target’s network using tools like \textit{Ping} and \textit{Fping} for more detailed sweeps.
 (4) Vulnerability Assessment: The pen-tester scans the identified systems for vulnerabilities using tools like Nessus, which compares the findings against a constantly updated vulnerability database.
 (5) Exploitation: The pen-tester attempts to exploit the discovered vulnerabilities to gain unauthorized access or escalate privileges. Exploits may target web applications (e.g., using OWASP tools) or systems (e.g., malware, and password attacks like brute force and dictionary attacks).
 (6) Reporting: The final phase involves delivering a detailed report that outlines detected vulnerabilities, associated risks, and recommendations. AI and ML can enhance this process by providing tailored insights based on the gathered data.

\subsection{Reinforcement Learning}
Reinforcement learning \citep{sutton_reinforcement_2018} constitutes a family of machine learning algorithms designed to solve problems modeled as Markov Decision Process (MDP).
An MDP provides a mathematical framework to describe the interaction between an agent and an environment. The agent aims to learn a policy $\pi$ that maximizes an objective function through interactions with the environment. Formally, the most basic MDP is defined as a tuple:
\begin{equation*}
    \langle \mathcal{S}, \mathcal{A}, \mathcal{T}, \mathcal{R} \rangle
\end{equation*}
where $\mathcal{S}$ is the set of states an environment can be represented, $\mathcal{A}$ is the set of actions the agent policy can choose to perform on the environment, $\mathcal{T}: \mathcal{S} \times \mathcal{A} \rightarrow \mathcal{S}$ is a probabilistic transition function defining the environment behavior evolution from one state to the next state upon an action selected by the agent's policy, and $\mathcal{R}: \mathcal{S} \times \mathcal{A}\times \mathcal{S} \rightarrow \mathbb{R}$ is a probabilistic reward function returning a single real-value scalar after taking an action in a certain state and arriving to a new state. 

For example, if we take Chess as an RL environment \citep{silver_mastering_2017}, a state $s\in \mathcal{S}$ represents the current position of the pieces on the board, an action $a\in \mathcal{A}$ corresponds to a legal move of a piece, and the reward $r\in \{-1,0,1\}$ is positive when the player wins, negative when they lose and zero otherwise. In section \ref{sec:environment}, we explain our choice of action and state space for the case of cybersecurity.
A key assumption when using MDPs is the Markov Property, which states that the system's dynamics depend only on the current state and not on the history. Although most environments in cybersecurity do not strictly satisfy this property, we can address this issue by incorporating a historical record of previous executions into the state representation. Another special case is whenever the environment is not fully observable and for each step, an observation $o\in \mathcal{O}$ is used instead. This slightly modified framework is named after the Partially Observable Markov Decision Process (POMDP). It is the most general way to encompass most real-world environments. We will use indistinctively the observation and the state although be aware of the difference. 
\\
\\
In reinforcement learning, a \textit{step} is the process of taking a single action and waiting for a response from the environment in terms of new action and reward. The sequence of steps is called an \textit{episode}.
An episode starts with an initial state $s_0\in \mathcal{S}$ that is either deterministic or given by a probability distribution, denoted as $s_0\sim p_0(\mathcal{S})$. It is common practice to start a pentesting attack using a single initial URL from the website whose vulnerabilities we want to analyze. Next, the policy chooses an action $a_0\sim \pi(\cdot|s_0)$ that can be performed in the state $s_0$. Then, the transition probability function returns the next state $s_1\sim \mathcal{T}(s_0, a_0)$. This new state contains information about the website that was obtained while performing the action $a_0$. Finally, the reward function returns a numerical value to measure the success of the performed action $r_1 = \mathcal{R}(s_0, a_0, s_1)$. This reward depends on the new information and vulnerabilities discovered in $s_1$. Moreover, a small quantity is subtracted from the reward to account for the cost of performing the action $a_0$, as some actions may be more computationally costly than others. A complete episode is given by
\begin{equation*}
    \begin{cases}
        a_t\sim \pi(\cdot, s_t)\\
        s_{t+1}\sim \mathcal{T}(s_t, a_t)\\
        r_{t+1}\sim \mathcal{R}(s_t, a_t, s_{t+1})
    \end{cases}
    \quad\quad\quad t=0,1,2,\dots
\end{equation*}
This process is repeated until we reach the maximum number of steps or until we have obtained every vulnerability. In practice, we do not know the number of vulnerabilities in a website; however, we do know the number of vulnerabilities in the training and validation environments.
\\
\\
During training, the Reinforcement learning algorithm uses the MDP framework to learn the optimal action policy $a \sim\pi^*(\cdot|s)$, which determines a distribution over actions $a\in \mathcal{A}$ given a state $s\in \mathcal{S}$, such that the expected cumulative reward, i.e., long-term reward, is maximized:
\begin{equation} \label{eq:optimization}
    \pi^* = \text{arg }\max_{\pi}G_t \approx  \text{arg }\max_\pi \sum_{t=t_i}^\infty \gamma^t \mathbb{E}_\pi[r_t],
\end{equation}
where $\gamma$ is the discount factor that is introduced for computational reasons, $\mathbb{E}_\pi [\cdot]$ represents the expected value with respect to the distribution given by the policy $\pi$, and $r_t$ is the reward obtained at time-step $t$. 
\\
\\
Depending on the degree of knowledge of the transition function $\mathcal{T}$, reinforcement learning algorithms can be classified as either model-based or model-free. Model-based algorithms assume that an explicit formula for the transition function is known and use it to obtain better policies. Model-free algorithms, on the other hand, do not have any prior knowledge about the environment or its dynamics. The cybersecurity realm is better described by model-free approaches, as we have no prior knowledge about the environment or its transition function; therefore, we must focus on these algorithms. However, we do not rule out the use of model-based algorithms in the future, as we are intrigued by their potentially superior performance.

\section{Environment}\label{sec:environment}

The most important aspect of Reinforcement Learning is correctly defining the environment and the reward function. The objective function in penetration testing is often based on achieving clear goals, such as \textit{Capture the Flag}. However, encoding this prior knowledge into the agent's policy and training this behavior is not straightforward. Instead, we define our intended goal in terms of the number and severity of vulnerabilities discovered in the analyzed website. This alternative reward model resembles a \textit{bug bounty} incentive and simplifies the problem formulation.
\\
\\
Cybersecurity practitioners utilize multiple platforms for vulnerability testing. These include public bug bounty platforms\footnote{Popular platforms for bug bounty programs include HackerOne (\url{https://www.hackerone.com/}), Bugcrowd (\url{https://www.bugcrowd.com/}), Synack (\url{https://www.synack.com/}), and Open Bug Bounty (\url{https://www.openbugbounty.org/}).}, Self-hosed bug bounty programs\footnote{Examples of self-hosted bug bounty programs include GitHub Security (\url{https://github.com/security}), Google's Vulnerability Reward Program (\url{https://bughunters.google.com/}), and Meta's Bug Bounty Program (\url{https://bugbounty.meta.com/}).}, Testing environments and labs\footnote{For hands-on practice and security training, see OWASP Juice Shop: \url{https://demo.owasp-juice.shop/}; HackTheBox: \url{https://www.hackthebox.com/}; and TryHackMe: \url{https://tryhackme.com/}.}, Virtual Machines and containers\footnote{Examples of popular virtual machines and containers include Metasploitable (\url{https://github.com/rapid7/metasploitable3}), DockerLabs (\url{https://dockerlabs.es/}) and Damn Vulnerable Web Application (DVWA) (\url{https://github.com/digininja/DVWA}).}. While these platforms are widely used for learning and skill development, many are extremely challenging and unsuitable for training policies from scratch. After careful review, we selected \textit{Damn Vulnerable Web Application} (DVWA) and DockerLabs due to its diverse range of vulnerabilities and configurable difficulty levels that enable progressive learning, potentially accelerate training, and provide consistent reward signals during the training/validation process. 
\\
\\
Several challenges persist regarding the environment. Despite the abundance of available platforms, few are compatible with our approach. A primary concern is that the agent might not generalize effectively and could overfit to a limited set of vulnerabilities. Additional drawbacks of using real environments include limited parallelization capabilities and slow execution speeds. Consequently, we opted to use simulated environments for training and validation of the RL agent, reserving the real environment solely for testing as a measure of success and generalization capability. This approach has been widely adopted in the literature as it offers plentiful benefits \citep{schwartz_autonomous_2019, ghanem_reinforcement_2018, lopez-montero_towards_2023, nguyen_pengym_2025}.

\subsection{Action space $\mathcal{A}$}

\begin{figure}[!tb]
  \centering
  \includegraphics[width=0.9\textwidth]{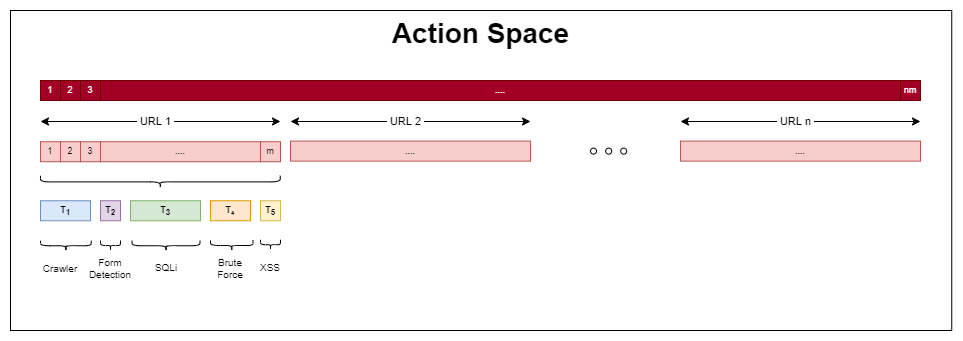}
  \caption{Action Space is composed of individual actions that can take place in each URL.}
  \label{fig:action_space}
\end{figure}
The true action space available in penetration testing is overwhelmingly large. Solving this task from scratch is intractable for modern AI due to the increasing complexity, requirement for long-term planning, and lack of available public datasets. One way to reduce the action space size is to use existing tools that can handle specific aspects of cybersecurity and focus on tool selection and configuration instead of low-level testing. We will denote henceforward, the set of consecutive natural numbers up to $n$ as follows $[n] := \{1,2,3,...,n\}$. 
The core tools that the agent can use focus mainly on OWASP Top 10 list\footnote{OWASP Top Ten: \url{https://owasp.org/www-project-top-ten/}.} and we have divided them into 5 main categories:
\paragraph{1. Crawler}
 Analyzes the website and searches for internal links recursively, mapping the URLs into a network graph. This tool has different configurations that the agent must choose from: (1) The \textit{relative depth} has 4 different levels, which represent the depth of the search starting from the node where the action is executed; (2) The brute force file offers 7 different files that contain lists of keywords used to discover untraceable endpoints. We use a collection of lists found in  \textit{SecLists}\footnote{\label{footnote:seclist}SecLists is a well-known catalog of lists used in cybersecurity.  GitHub: \url{https://github.com/danielmiessler/SecLists}}). Therefore, the set of actions related to this task can be calculated as $T_1:=\{1,2,3,4\} \times \{1,2,3,...,7\}$, hence, the total number of actions is $|T_1| = 28$.
\paragraph{2. Form and Parameter Detection} This action uses a web-scraping tool to extract forms and parameters that are later used to inject and brute-force other vulnerabilities. This tool has no configuration options, and therefore $|T_2|=1$.
\paragraph{3. SQL Injection (SQLi)} Uses built-in tools that incorporate fuzzing commands to inject SQL code into previously discovered parameters. There are multiple configuration parameters that the model can utilize: (1) The \textit{level} number specifies the completeness of the tests to be performed. The value ranges from 1 to 5. The higher the value, the greater the number of payloads tested; (2) The \textit{risk} specifies a filter on the payloads used. The greater the number, the riskier the action that is performed. This value ranges from 1 to 3; (3) There are 6 different techniques that the agent can choose to exploit: boolean-based blind, time-based blind, error-based, UNION query-based, and stacked queries (note: this lists only 5 techniques); and (4) Given the forms and parameters discovered in the URL, the agent must choose which form to use to inject the code. It depends on the maximum number of forms allowed; for simplicity, we assume there is only one, although in practice, the model will take into account all available information. Therefore, $T_3:=\{1,2,3,4,5\} \times \{1,2,3\} \times \{1,2,3,4,5,6\}$, hence, $|T_3|=90$.
\paragraph{4. Brute Force Techniques}: Uses a list of common usernames and passwords from \textit{SecLists}\textsuperscript{\ref{footnote:seclist}}. The agent can choose from a list of 4 different usernames and 6 different passwords. Hence, the action space is given by $T_4 = \{1,2,3,4\} \times \{1,2,3,4,5,6\}$ and $|T_4| = 24$.
\paragraph{5. Cross-Site Scripting}: Similar to SQLi, the agent selects from a configuration matrix depending on different levels of complexity resulting from a tool that contains multiple payloads. The action space is described by $T_5 = \{1,2,3\}$ with $|T_5| = 3$.

The action space available for one URL, denoted by $\mathcal{A}_{url}$, is given by $\mathcal{A}_{url} := T_1 \sqcup T_2 \sqcup T_3 \sqcup T_4 \sqcup T_5$, where the operator $\sqcup$ denotes the disjointed union of sets. Therefore, the total number of actions that can be executed in a single URL is given by $m:=|\mathcal{A}_{url}| = \sum_{i=1}^5 |T_i| = 134$. Each action encompassed in $\mathcal{A}_{url}$ can be used in different nodes/URLs of the website, we denote the number of URLs found as $n$. The total action space is the Cartesian product of the number of URLs by the number of actions in each URL, i.e., $\mathcal{A} := [n] \times \mathcal{A}_{url}$. In other words, the total number of actions in the action space is given by $|\mathcal{A}|=134n=mn$. Note that the action space is dynamic and depends on the number of URLs already discovered in the application. The action space grows as the agent progresses and discovers more information and vulnerabilities.

\subsection{State space $\mathcal{S}$}
We define the \textit{observation} $\mathcal{O}$ as the information encompassing the set of actions performed and their corresponding responses in previous time steps. In other words, if $o_t$ and $o_{t+1}$ are the observations at time steps $t$ and $t+1$, respectively, then the information contained in $o_{t+1}$ always supersedes that in $o_t$. The observation is updated as new information becomes available and is used to determine the reward of an action.
We denote the newly acquired information resulting from executing action $a_t$ as $o_{t+1} \setminus o_t$. This formulation allows us to isolate the reward obtained by each action, enhancing the model's ability to learn the causal attribution of actions.
Since the observation consists of unstructured data, it cannot be directly used as input for a machine-learning-based architecture. To address this, we have designed a function that encodes the observation into a structured format suitable for the model, namely the state space:
\begin{align*}
    \ttt{encode}:\; &\mathcal{O}\longrightarrow\mathcal{S}\\
    &o_t \longmapsto s_t
\end{align*}
Henceforth, we will consider the observation and state space as interchangeable concepts to simplify notation, with the distinction between the two being clear from the context. 

The temporal component is crucial in cybersecurity, where the sequence of actions significantly impacts outcomes. Although recurrent networks would be suitable for handling this temporal dimension, we opted against them due to the already extensive action space. Instead, we encoded the temporal component using a cumulative process with exponential decay. This temporal encoding works by applying a decaying factor to previously observed information. Specifically, given a piece of information in the state space $s_i\in\mathbb{R}$ at time-step zero, in the absence of other observation during the next $t$ time-steps, the value of its observation decays to $s_i\lambda^t$ where $\lambda\in (0,1)$ is the decay factor. This approach allows us to retain historical information while gradually reducing its influence over time.
\\
\\
To track when an action has been performed and its outcome, we use a state representation where zero indicates an action has not been previously executed, and for actions that have been executed, we record the latest reward $r_t$ obtained. When an action is executed multiple times, the state reflects only the most recent result. This approach creates a state space with dimensionality equal to the action space, i.e., $|\mathcal{S}|=|\mathcal{A}|$. However, our tools extract additional information that enhances decision-making, such as URL status and tool versions. Let us define the basic state for a single URL with dimensionality matching its action space counterpart: $\mathcal{S}^*_{url} := \mathbb{R}^{|\mathcal{A}_{url}|}$. We then augment this with additional features as a vector of dimension $n_f$.Therefore, the complete state space for a single URL is $\mathcal{S}_{url} := \mathcal{S}^*_{url} \times \mathbb{R}^{n_f} = \mathbb{R}^{|\mathcal{A}_{url}| + n_f}$. Finally, the total state space is constructed as the product of individual URL state spaces across all $n$ URLs: $\mathcal{S}:=\mathcal{S}_{url} \times \overset{n}\cdots\times S_{url}$. 

\subsection{Reward $\mathcal{R}$}

Correctly modeling the reward function is fundamental; well-designed incentives accelerate learning and reduce volatility during training. The overarching goal is to maximize the discovery of critical vulnerabilities in the shortest possible time. Thus, we formulate the problem as a bi-objective optimization task:
\begin{align*}
    \mathcal{R}: \mathcal{S}\times \mathcal{A}\times \mathcal{S} &\longrightarrow \mathbb{R}\\
    (s_{t}, a_{t}, s_{t+1}) &\longmapsto \mu \mathcal{V}(s_t, s_{t+1}) + (1-\mu)\mathcal{C}(a_t)
\end{align*}
where $\mathcal{V}(s_t, s_{t+1})$ is a positive reward for discovering new information or vulnerabilities at time step $t$ (i.e., comparing the previous state $s_t$ with the updated state $s_{t+1}$), and $\mathcal{C}(a_t)$ represents the negative cost associated with the computational resources and time consumed by action $a_t$. The parameter $\mu \in (0,1)$ controls the trade-off between these objectives. We frame the problem as in (\ref{eq:optimization}) and solve it using reinforcement learning. 
\\
\\
The positive reward assigned by $\mathcal{V}_t$ is proportional to the criticality of the newly discovered vulnerabilities. A more detailed breakdown of the positive reward, based on a wide range of information, is presented in Table \ref{tab:positive_reward} (Appendinx \ref{appendix:1}). The action cost denoted by $\mathcal{C}_t$, is estimated based on computational time and associated risk, and it remains fixed regardless of the action's outcome. Table \ref{tab:cost} provides a comprehensive overview of the costs associated with different actions. These values are carefully fine-tuned and incorporated into the reward model. The agent must balance reward and cost, as higher-reward vulnerabilities typically require higher-cost actions.

\subsection{Simulated Environment}
\begin{figure}[!htb]
  \centering
  \includegraphics[width=0.9\textwidth]{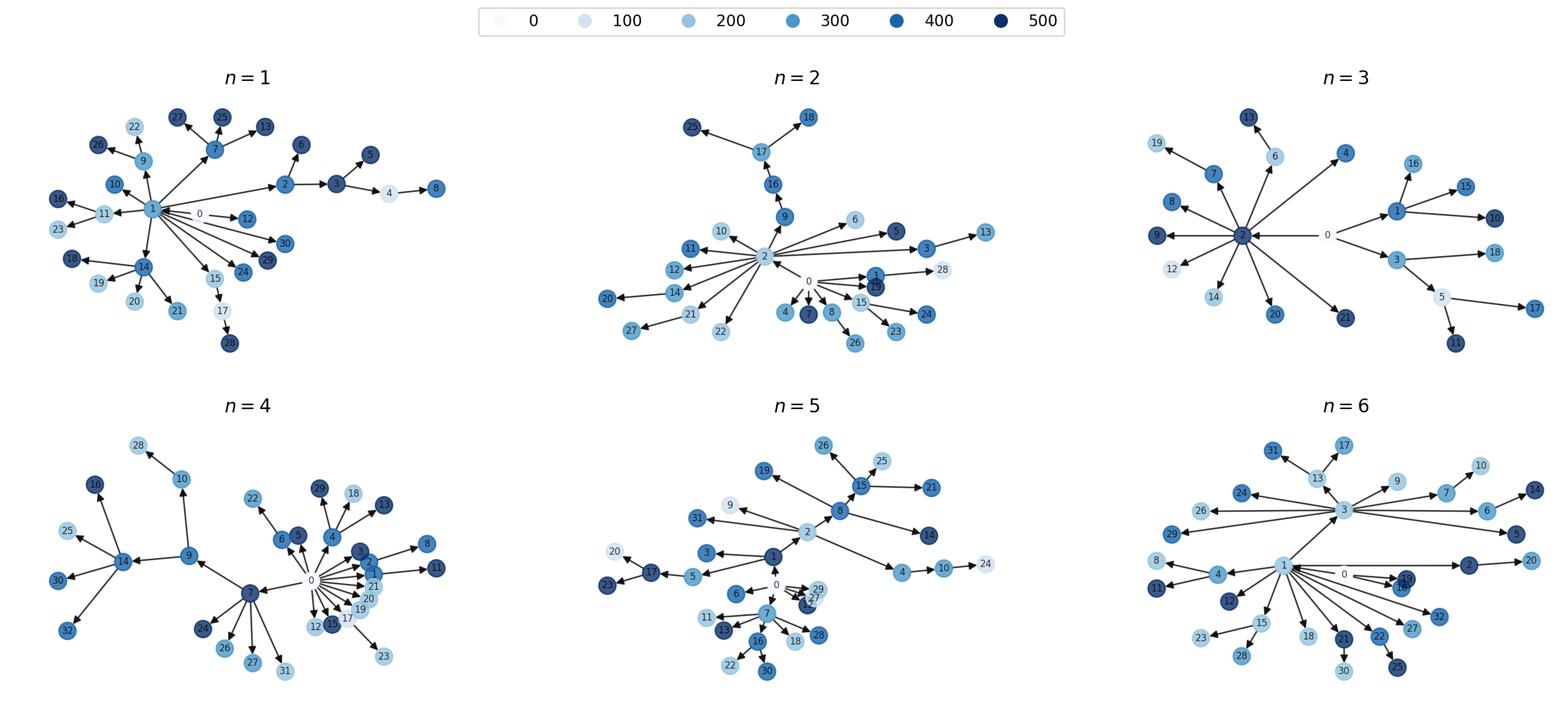}
  \caption{Six randomly generated environments used for training and agent validation. The colors indicate the status code of each URL.}
  \label{fig:graph}
\end{figure}

The simulated environments are created from probability distributions, allowing us to use them for training and validation without concerns about overfitting. Additionally, these environments enable significantly faster execution compared to real-world settings, achieving nearly a 3000x speedup, and support parallelized action execution \citep{nguyen_pengym_2025, schwartz_pomdp_2020, emerson_cyborg_2024, carrasco_cybershield_2024}. However, these advantages come with trade-offs. Simulated environments may introduce biases toward certain vulnerabilities and may not accurately reflect the true distribution of real websites. Furthermore, each action requires creating a simulation that closely mimics the behavior of real-world tools, which can be a labor-intensive process. Despite these challenges, the benefits far outweigh the costs.
\\
\\
The simulation environment maintains an internal state used to generate and evaluate the outcomes of actions, and it is procedurally generated at random. First, we construct the website’s topology. Like any real website, it has a root node from which URL nodes branch out, forming a tree structure—meaning the graph contains no cliques. To generate this random tree network, we use a modified version of the Barabási-Albert algorithm \citep{barabasi_emergence_1999} (see Algorithm \ref{alg:random_tree}). This algorithm ensures that the resulting graph remains a tree while mimicking real-world website structures by preferentially adding edges to nodes with higher degrees. The resulting topology can be observed in Figure \ref{fig:graph}.

\begin{algorithm}[!tb]
\caption{Modified Barabási-Albert Algorithm}\label{alg:random_tree}
\begin{algorithmic}
\Require Number of nodes $n\sim \text{Poisson}(\lambda=40)$.\\
Initialize graph $\mathscr{G} = (\mathscr{V}, \mathscr{E})$ where $\mathscr{V} = \{1,2,\dots, n\}$ and $\mathscr{E} = \{(1,2)\}$. \\
Let $M := \{1,2\}$ be a multiset.
\For{$2<i\leq n$}
\State \texttt{source} $\leftarrow$ Uniform sampling from $M$.
\State $\mathscr{E} \leftarrow \mathscr{E} \cup \{(\text{\texttt{source}}, i)\}$
\State $M \leftarrow M \cup \{\text{\texttt{source}}, i\}$
\EndFor
\end{algorithmic}
\end{algorithm}

Once the website topology has been populated, the next step is to initialize the nodes with vulnerabilities and essential information required for executing actions. Each node samples from random variables, and these variables are independent across nodes. Specifically, each node draws from a distribution to determine the status code for its assigned URL. For every valid status code, we assign tools and versions based on a probability distribution and introduce vulnerabilities similar to those in Table \ref{tab:positive_reward}. 
Beyond defining vulnerabilities and their types for each URL, we also configure the tools and their actions to enable the exploitation of these vulnerabilities. For instance, if a node includes an SQL injection vulnerability, we might configure a Boolean-based SQL injection with a minimum level of 3 and a risk value of 1. This ensures that the vulnerability can only be exploited if a Boolean-based action is selected with a risk level and severity meeting or exceeding 3 and 1, respectively.
The behavior of the simulated actions must mirror the real action's internal logic, for each action, we emulate a simplification of this process and are validated to fit a general purpose use of the tools. Note that the state space and action space must remain the same in order to use the agent in a real-world scenario afterwards.

\section{Agent and Geometrical Deep Learning}\label{sec:agent}

The agent employs a model-free policy, reflecting real-world scenarios. We experiment with various on-policy and off-policy algorithms, including SAC \citep{haarnoja_soft_2019}, PPO \citep{schulman_proximal_2017}, and using DQN \citep{mnih_playing_2013} as a benchmark. SAC and PPO follow a hybrid approach combining two components: (1) the \ttt{Actor} which selects actions based on a learned policy, and (2) the \ttt{Critic}, which evaluates actions by estimating the value function and providing feedback. We detail the agent's mathematical implementation and focus on PPO, as it yields the best results, though SAC and DQN share similarities. The actor and critic function is given by:
\begin{equation}
    \begin{aligned}
    \ttt{Actor}:\;& \mathcal{S} \rightarrow \mathbb{R}^{|\mathcal{A}|} = \mathbb{R}^{|\mathcal{A}_{url}| \times n}\\
    \ttt{Critic}:\,& \mathcal{S}\rightarrow\mathbb{R}.
    \end{aligned}
\end{equation}
The \texttt{Actor} outputs logits that determine the probability of selecting each action. These probabilities are computed using functions like softmax to ensure they sum to 1. The \texttt{Critic} network evaluates the current state, i.e., $V(S)$. A conventional Multi-Layer Perceptron is impractical due to the vast and dynamically changing state and action spaces. For example, in a website network of $n$ URLs ($n\geq 50$) we have $|\mathcal{A}| = n\times |\mathcal{A}_{url}| \geq 6700$ and $|\mathcal{S}| = n\times |S_{url}| \geq 6700$. A fully-connected 1-layer neural network, we would have more than 36M parameters. Even with hidden layers, the large action space leads to high complexity.
To address this, we exploit problem symmetries. Notably, the \texttt{Critic} network should be invariant to the order of URLs, a property that can be expressed mathematically as follows:

\begin{equation}
    \ttt{Critic}((s_1,s_2,\dots, s_{n})) = \ttt{Critic}((s_{\sigma(1)},s_{\sigma(2)},\dots, s_{\sigma(n)}))\quad\quad \forall \sigma \in \text{Sym}(n)\quad  \forall s_1,s_2,\dots, s_{n} \in \mathcal{S}_{url}
\end{equation}

where Sym($n$) is the group of permutations (or symmetries) of $n$ elements. This property is called \textit{permutation invariance} because it remains unchanged under any permutation of the states. Note that the composition of functions that are permutation invariant is also invariant. Similarly, the \texttt{Actor} network should produce a correspondingly permuted output when the order of the nodes is altered. Assume that

\begin{equation*}
    (p_1,p_2,\dots, p_{n}):=\ttt{Actor}((s_1,s_2,\dots, s_{n}))
\end{equation*}

where $p_i \in \mathbb{R}^{|\mathcal{A}_{url}|}$ are the models' output logits. Then, the network should satisfy the following permutation symmetry:

\begin{equation}
    \ttt{Actor}((s_{\sigma(1)},s_{\sigma(2)},\dots, s_{\sigma(n)})) = (p_{\sigma(1)},p_{\sigma(2)},\dots, p_{\sigma(n)}) \quad\quad \forall \sigma \in \text{Sym}(n)\quad  \forall s_1,s_2,\dots, s_{n} \in \mathcal{S}_{url}
\end{equation}

When a function has this property, it is called \textit{permutation equivariance} because permuting the input also permutes the output. These properties are extensively studied in Geometric Deep Learning \citep{bronstein_geometric_2021}. Establishing such geometric priors reduces the search space dimensionality and mitigates the curse of dimensionality. Since a standard MLP does not inherently include this prior, we must adopt a different approach. Below, we detail the network architecture we have chosen.

Regarding the \texttt{Critic}, we define the following MLP over a partition of the real state space:
\begin{equation}
    \ttt{MLP}_C: \mathcal{S}_{url}\rightarrow \mathbb{R}
\end{equation}

We then apply an aggregation function \texttt{Aggr} that verifies the permutation invariance property. In our case, we use the standard summation which preserves this property. Moreover, the summation aligns with the Critic's goal of assessing the overall state value. Hence, we can define:
\begin{equation}
\begin{aligned}
    \ttt{Critic}: \mathcal{S}_{url} \times \overset{n}{\dots}\times \mathcal{S}_{url} &\longrightarrow \mathbb{R}\\
    (s_1, s_2, \dots, s_n)& \longmapsto \ttt{Aggr}(\ttt{MLP}_C(s_1), \ttt{MLP}_C (s_2), \dots, \ttt{MLP}_C (s_n)) = \sum_{i=1}^n \ttt{MLP}_C (s_i)
\end{aligned}
\end{equation}

This formulation reduces the number of parameters and allows parallel computation.
For the \texttt{Actor} network, we proceed similarly, defining the following parametrized function:
\begin{equation}
    \ttt{MLP}_A: \mathcal{S}_{url} \rightarrow \mathbb{R}^{|\mathcal{A}_{url}|} 
\end{equation}
The \texttt{Actor} is then obtained by concatenating the results for different URLs:

\begin{equation}
\begin{aligned}
    \ttt{Actor}: \mathcal{S}_{url} \times \overset{n}{\dots}\times \mathcal{S}_{url} &\longrightarrow \mathbb{R}^{|\mathcal{A}_{url}|} \times \overset{n}{\dots}\times \mathbb{R}^{|\mathcal{A}_{url}|}\\
    (s_1, s_2, \dots, s_n)& \longmapsto \ttt{MLP}_A(s_1)  \| \dots \| \ttt{MLP}_A(s_n)
\end{aligned}
\end{equation}

where $\|$ denotes concatenation. This function satisfies permutation equivariance; however, it assumes that the state of one URL does not influence the probability of an action in another. While this assumption may not always hold, there is no indication that it negatively impacts model performance.

From the perspective of geometric deep learning, both the \texttt{Actor} and \texttt{Critic} can be viewed as a special case of a Graph Neural Network, where the graph consists of isolated nodes (URLs) with no edges.

\section{Experiments}\label{sec:experiments}

\begin{table}[!b]
\centering
\begin{tabular}{lcc}
\hline
\textbf{Hyperparameter} & \textbf{Importance} & \textbf{Best Value} \\
\hline
RL algorithm & \textbf{0.67} & PPO \\
Steps per episode & 0.11 & 66 \\
Arquitecture & 0.10 & [64, 32] \\
Initial lr & 0.08 & $3.29\times 10^{-3} $\\
Batch size & 0.04 & 256 \\
\hline
\end{tabular}
\vspace{4pt}
\caption{Hyperparameter importance analysis and optimal values for the reinforcement learning model}
\label{tab:hyperparameters}
\end{table}

\begin{figure}[!tb]
    \centering
    \includegraphics[width=0.8\linewidth]{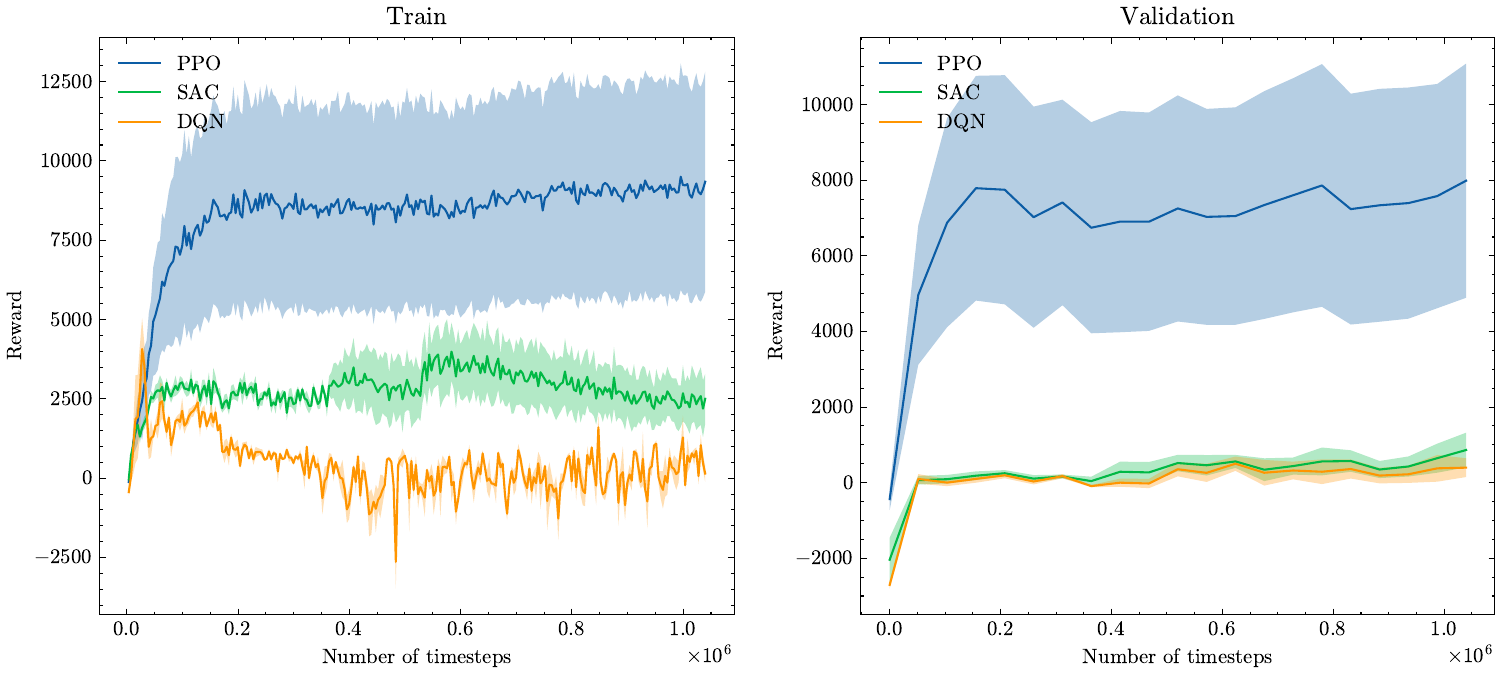}
    \caption{Training and validation during 1e6 timesteps.}
    \label{fig:train}
\end{figure}

\begin{figure}[!tb]
    \centering
    \subfigure[Number of actions performed on a single URL. The most common case is performing between 0 and 5 actions per URL. ]{\label{fig:url_proportion}\includegraphics[width=0.45\textwidth]{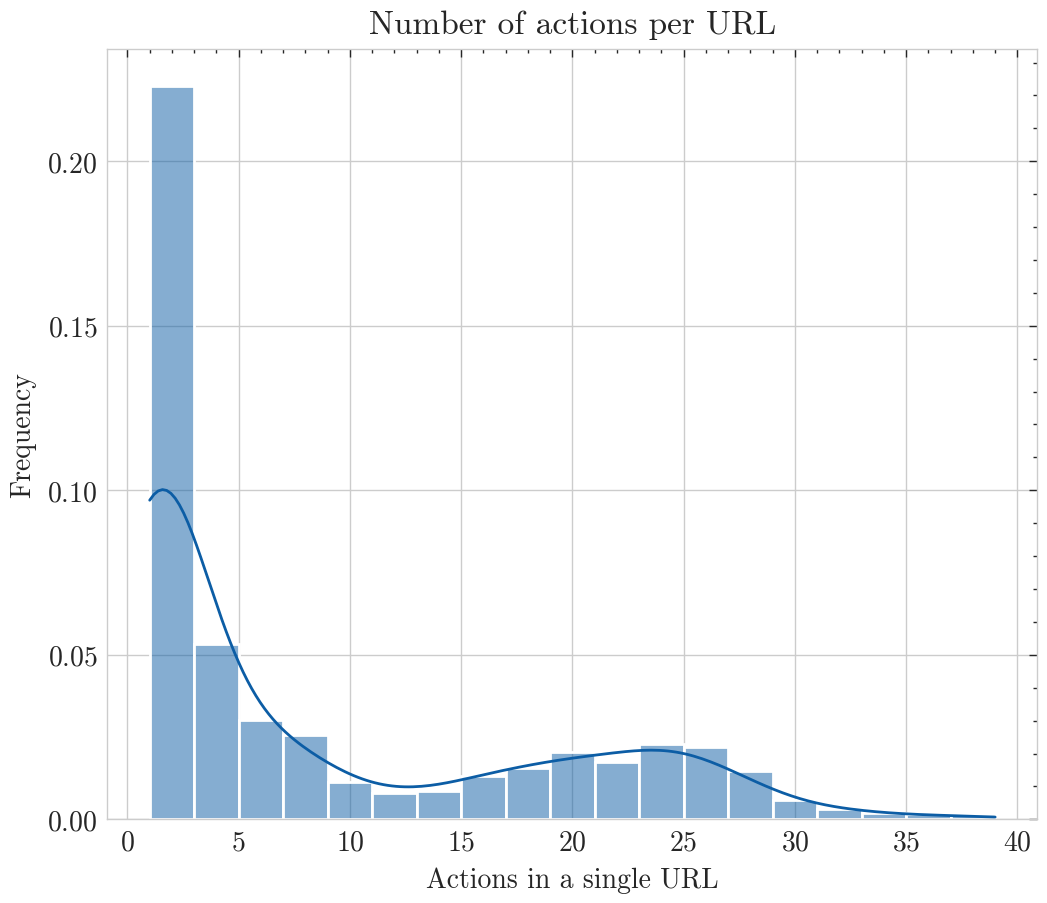}}
    \hfill
    \subfigure[Proportion of tool selections by the agent. It is unsurprising that SQLi was executed more frequently, given the higher number of configurations.]{\label{fig:action_proportion}\includegraphics[width=0.40\textwidth]{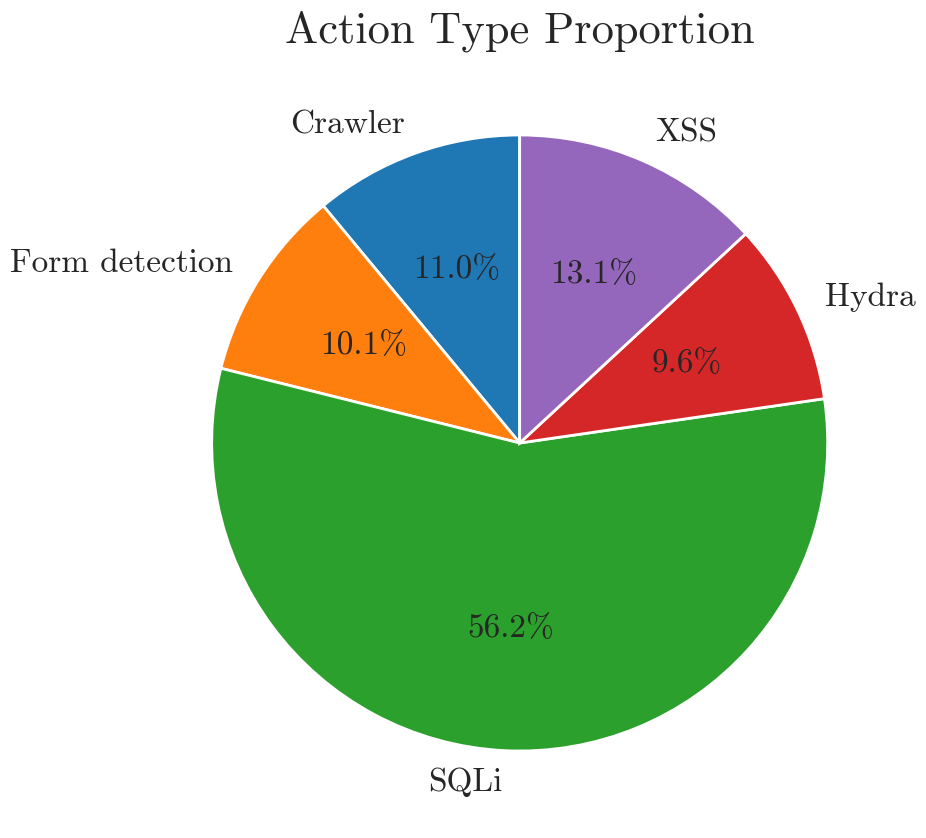}}
    \caption{Results obtained during testing at the end of each episode (500 timesteps).}
\end{figure}

\begin{figure}[!tb]
    \centering
    \includegraphics[width=0.9\linewidth]{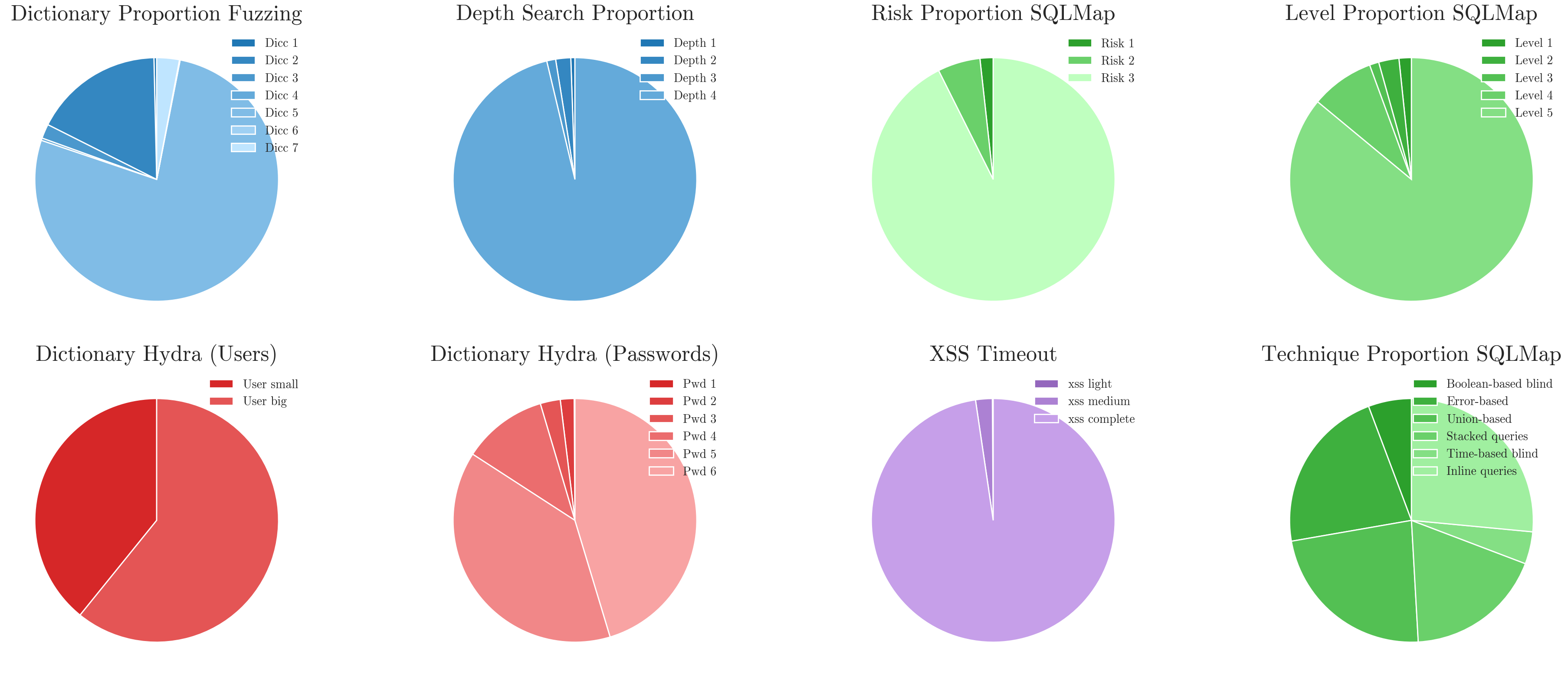}
    \caption{Sub-action selection probability during inference. Actions within the same tool are represented by the same colors.}
    \label{fig:metrics}
\end{figure}

We train the reinforcement learning (RL) model using samples from the simulation environment and evaluate the agent on the real website \textit{DVWA}\footnote{Damn Vulnerable Web Application (DVWA): \url{https://github.com/digininja/DVWA}}. and \textit{DockerLabs}\footnote{DockerLabs is hub that collects multiple vulnerable docker machines: \url{https://dockerlabs.es/}}. The training process was conducted on the following hardware: a Tesla T4 GPU with 15GB of RAM, an Intel Xeon Silver 4214 processor with 24 cores and 48 threads, and 256GB of RAM. Training for 1e6 timesteps takes between 3 to 8 hours, depending on the training configuration. While the process does not fully utilize the available RAM, it consumes approximately 40–60GB. Additionally, we observed GPU usage fluctuating between 30\% and 40\% during training.
\\
\\
To optimize the hyperparameters in the RL algorithm, we employ the TPESampler algorithm~\citep{ozaki_multiobjective_2022}, a Bayesian-based hyperparameter optimizer that utilizes a Gaussian Mixture Model (GMM) and is well-suited for computationally expensive optimization problems.
The key hyperparameters we aim to optimize include: (1) RL algorithm selection: choosing between PPO, SAC, and DQN; (2) Number of steps per episode; (3) network architecture: allowing the algorithm to determine the hidden layer size, given a two-layer network; (4) initial learning rate: using a linear scheduler; (5) batch size: selecting from different powers of two. The hyperparameter optimization process is conducted over 500k timesteps to enable faster iterations. We evaluate 300 different configurations, sampling 10 environments for training and 10 for validation. Performance is assessed on the validation set to derive a numerical score for comparison across different runs. The best-performing configuration is presented in Table \ref{tab:hyperparameters}.
\\
\\
We train a model using the three mentioned algorithms, employing the optimal configurations obtained from the hyperparameter search. The model is trained 1e6 timesteps, with 200 steps per episode, allowing for more complex behaviors and the development of chained strategies. It consists of 69,304 parameters—a reasonable size given the large action and observation space.
In total, we sample 60 environments from the distribution and divide them into training and validation sets, with 50 samples allocated for training and 10 for validation. Figure \ref{fig:train} compares the performance of the three algorithms. As expected, the on-policy algorithm learns faster than SAC and DQN. The PPO algorithm successfully identifies vulnerabilities, maximizes rewards, and generalizes well to the validation set.
\\
\\
The testing was conducted on multiple machines, including \textit{DVWA} and \textit{DockerLabs}. The agent successfully extracted all vulnerabilities within its capabilities. However, some vulnerabilities remain unexploitable due to limitations in the tools available to the agent, making them infeasible. Certain results are logged during inference for further analysis. Figures \ref{fig:action_proportion} and \ref{fig:url_proportion} provide high-level metrics on the model's performance, while Figure \ref{fig:metrics} presents detailed statistics on the proportion of each sub-action selected by the agent. In some cases, these insights were used to fine-tune the reward function and improve the simulation.
Although these statistics alone are insufficient for a comprehensive assessment, they help detect anomalies during inference and identify potential biases in the model. For example, the team leveraged them to refine the reward mechanisms and enhance the simulation's effectiveness.

\section{Results and Future Prospects}\label{sec:results}
In this paper, we introduce a novel approach to penetration testing for web applications, addressing the complexity of this inherently challenging task. While our method does not yet achieve professional human-level performance and remains constrained in its range of available actions compared to the vast possibilities, it marks a significant advancement. Unlike previous research, which often relied on controlled environments with limited real-world applicability, our approach improves upon these limitations. We train and validate our algorithms in diverse simulated environments and assess their effectiveness in real-world testing scenarios. By bridging the gap between simulation and practical deployment, our solution demonstrates promising results.

Our reinforcement learning (RL) model incorporates techniques inspired by geometric deep learning, leveraging inherent data symmetries to reduce the number of parameters required to map a wide range of observations and actions. Notably, our best-performing model operates with only 69,304 parameters. To circumvent the complexity of recurrent networks, we designed a structured observation space that inherently accounts for past rewards, effectively capturing temporal dependencies without the need for recurrence.

While our training process relies on model-free algorithms, the simulation framework provides opportunities to incorporate model-based techniques, potentially enhancing performance. Expanding the agent’s action set could further improve its capabilities. Additionally, there is significant room for improvement in how the agent processes and encodes information. Valuable insights are often embedded in string-based and unstructured data, such as HTML code. To address this, we are exploring the use of large language models (LLMs) to extract and encode this information into a more structured and manageable format.

\section*{Funding Statement}

This work has been primarily funded by the Spanish National Cybersecurity Institute (INCIBE), under the Spanish Ministry of Economic Affairs and Digital Transformation (Project ISOC, Exp. CPP2\_R29). We appreciate the financial support of GMV and its ongoing commitment to research and development. 

\section*{Contributions}

Investigation, D. L-M, J. A-A, A. M-M; mathematical approach, D. L-M; software implementation, D. L-M, J. A-A, A. L-M; model training, D. L-M, J. A-A, A. M-M; manuscript D. L-M, J.M. A-G; project management M. G-L, J.M-A; use case, business understanding M. G-L, J.M. M-A.

\section*{Acknowledgments}
We would like to express our sincere gratitude to GMV and our colleagues for providing invaluable support and collaboration throughout this research endeavor. 

\section*{Ethics declarations}
\subsection*{Conflict of interest}
The authors declared that they have no conflict of interest in this work.


\bibliographystyle{plainnat}  

\bibliography{references.bib}

\newpage
\appendix

\section{Report Generation}

Generating a penetration testing (pentesting) report involves documenting the findings, methodologies, and recommendations from a penetration test in a structured format. A well-crafted report typically includes:
\begin{itemize}\setlength\itemsep{0.15em}
    \item \textbf{Executive Summary} – A high-level overview for non-technical stakeholders.
    \item \textbf{Technical Details} – A comprehensive section outlining identified vulnerabilities, their severity, potential impacts, and recommended remediation steps.
\end{itemize}
The objective is to provide actionable insights that strengthen the organization's security posture while ensuring clarity, accuracy, and relevance for both technical teams and decision-makers.

To streamline this process, we have developed a script that automatically generates reports based on the findings obtained by the testing agent. These reports include supporting evidence that verifies each vulnerability. Additionally, every identified vulnerability is mapped to a Common Vulnerabilities and Exposures (CVE) entry—a publicly accessible database of known security flaws (NVDLib\footnote{NIST National Vulnerability Database: \url{https://nvdlib.com/en/latest/}}). Each vulnerability is also rated based on its criticality, allowing organizations to prioritize remediation effectively.

\section{Interface}
\begin{figure}[!tb]
    \centering
    \subfigure[Web interface main page. ]{\includegraphics[width=0.4\textwidth]{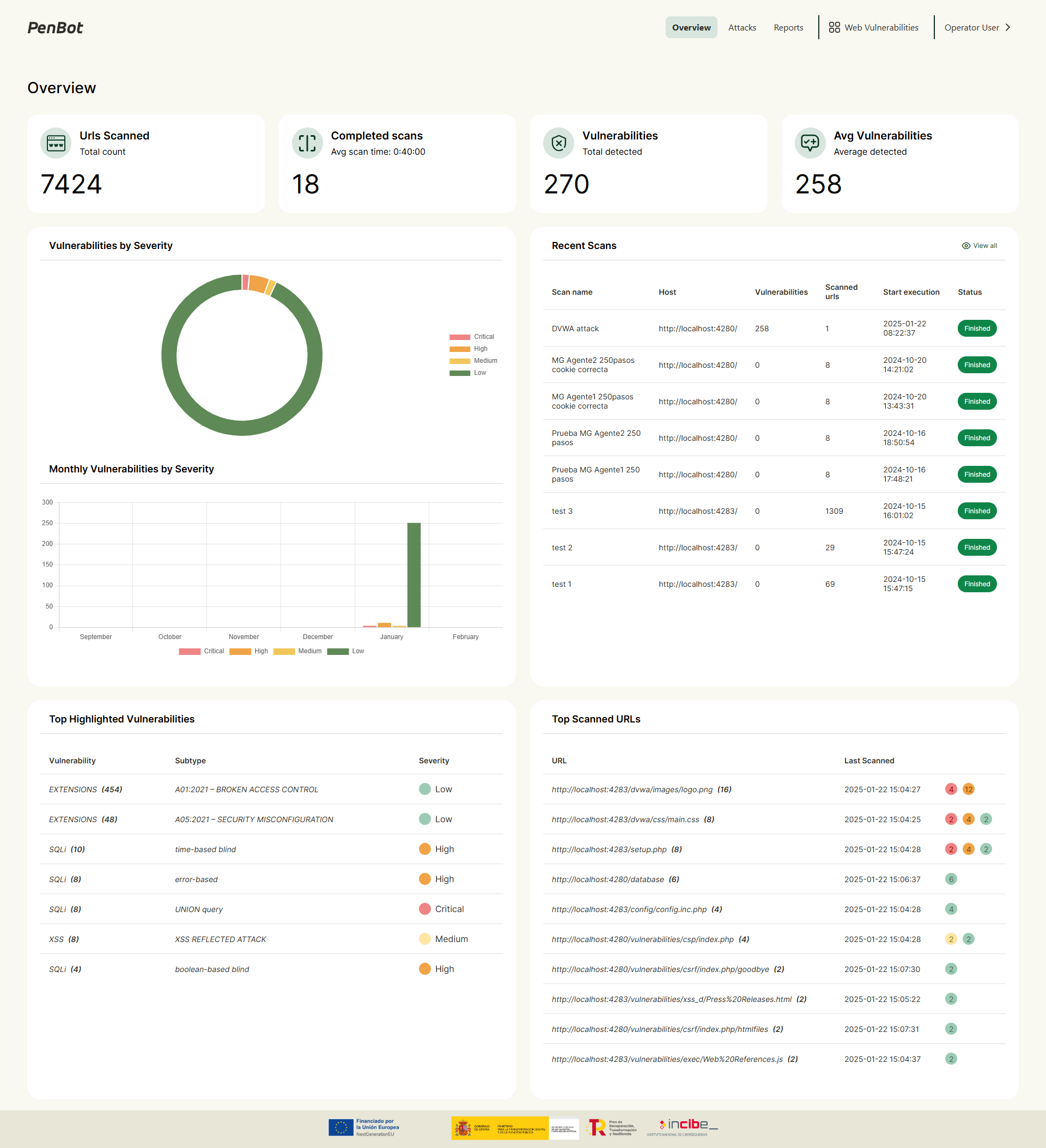}}\label{fig:penbot}
    \hfill
    \subfigure[Vulnerability page after each.]{\includegraphics[width=0.5\textwidth]{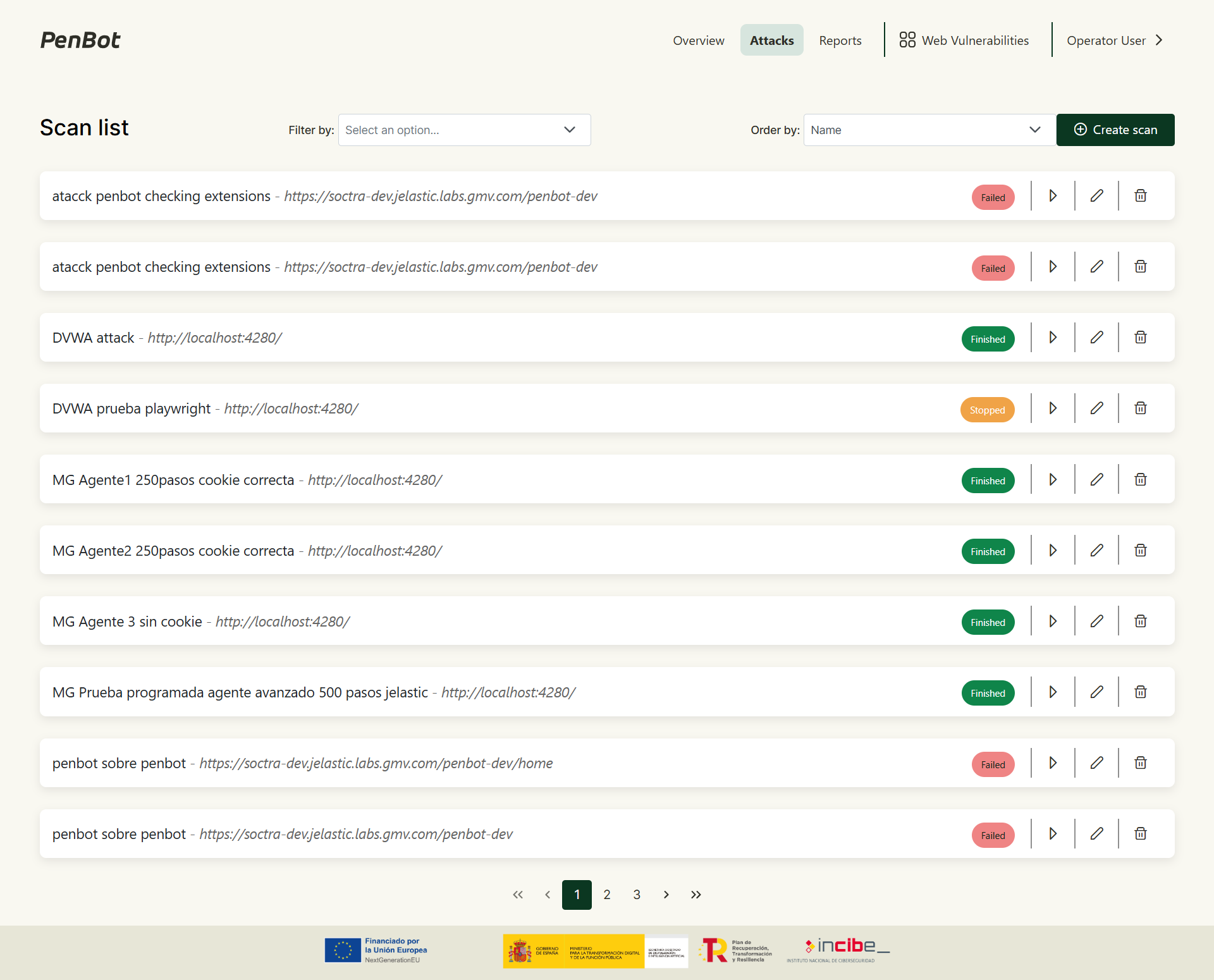}}\label{fig:penbot2}
    \caption{Penbot Web interface.}
    \label{fig:interface}
\end{figure}

We developed an interface that allows users to interact with the agent seamlessly (see Figure \ref{fig:interface}). It provides essential functionalities tailored to the end user’s needs. Among its key features is the ability to execute and schedule attacks using two different agents:
(1) \textbf{Advanced Agent} – Incorporates all available actions and can perform computationally intensive tasks with extensive configurations.
(2) \textbf{Basic Agent} – Optimized for speed, focusing on quick and shallow vulnerability assessments.

Once an attack is completed, the execution results are stored, and the interface displays relevant statistics along with a detailed report.
The application also includes a feature to review past executions and compare their performance, making it ideal for maintenance tasks and periodic website scans. Additionally, it offers a graphical representation of the website’s topology and allows users to upload logs of the tools utilized by the agent.

\section{Reward Definition}
\label{appendix:1}

We include the reward table used to train the agent in the simulation. The rewards and costs are categorized into different subtypes. The reward (Table \ref{tab:positive_reward}) depends on whether the agent successfully achieves its objective. In contrast, the cost (Table \ref{tab:cost}) represents a negative reward that applies to every action. The specific cost value is determined by the parameters set in the model. For instance, executing the crawler action at maximum depth incurs a higher cost—in terms of time and computation—compared to running it at minimum depth.
\begin{table}[!htb]

    \centering
    \scriptsize
    
    \begin{tabular}{llc p{0.55\linewidth}}
        Title & Subtype & Value & Description \\
        \midrule
         New Tools and Versions & &4 & The tools discovered and versions expose possible vulnerabilities. \\
         
         \arrayrulecolor{black!30}\cmidrule{1-4}
         
        \multirow{5}{*}{New URLs found} & Status [100,200) &1 & \multirow{5}{0.99\linewidth}{HTTP status codes are categorized into different brackets based on the type of response they represent: [100, 200) for informational responses indicating that the request is being processed; [200, 300) for successful responses where the request was received, understood, and accepted; [300, 400) for redirection, requiring further action by the client; [400, 500) for client errors, indicating issues with the request; and [500, 600) for server errors, where the server fails to fulfill a valid request due to internal problems.
} \\
        \arrayrulecolor{black!30}\cmidrule{2-3}
         & Status [200,300) &8 & \\
         \arrayrulecolor{black!30}\cmidrule{2-3}
         & Status [300,400) &6 & \\
         \arrayrulecolor{black!30}\cmidrule{2-3}
         & Status [400,500) &1 & \\
         \arrayrulecolor{black!30}\cmidrule{2-3}
         & Status [500,600) &1 & \\

        \arrayrulecolor{black!30}\cmidrule{1-4}
        Parameters found & &20 & The parameters found in forms can be used in other vulnerabilities, therefore we incentive effusively the discovery. \\
        
        \arrayrulecolor{black!30}\cmidrule{1-4}
        \multirow{5}{*}{SQL injection} & Boolean-based blind &60 & \multirow{5}{0.99\linewidth}{Boolean-based Blind and Time-based Blind methods are slower and less direct, requiring multiple requests to extract data, with moderate impact. Error-based techniques, which expose data through error messages, offer a quicker data leak and are more critical. UNION Query-based techniques allow direct data extraction, making them highly impactful and critical. Stacked Queries, the most severe, enable the execution of multiple commands in a single request, leading to potentially catastrophic outcomes.}\\
        \arrayrulecolor{black!30}\cmidrule{2-3}
         & Time-based blind &60 & \\
         \arrayrulecolor{black!30}\cmidrule{2-3}
         & Error-based &80 & \\
         \arrayrulecolor{black!30}\cmidrule{2-3}
         & UNION query-based &90 & \\
         \arrayrulecolor{black!30}\cmidrule{2-3}
         & Stacked queries &100 & \\
         \arrayrulecolor{black!30}\cmidrule{2-3}
         & Inline queries &100 & \\
         
        \arrayrulecolor{black!30}\cmidrule{1-4}
        \multirow{3}{*}{Cross-site Scripting} & Stored& 90 & \multirow{3}{0.99\linewidth}{Stored XSS is rated the highest due to its potential for widespread impact. DOM-Based XSS reflects its complexity and the subtlety of attacks. Reflected XSS is rated lower as it generally requires user interaction and is less persistent.} \\
        \arrayrulecolor{black!30}\cmidrule{2-3}
         & Reflected & 70 & \\

        \arrayrulecolor{black!30}\cmidrule{1-4}
         Password Brute Force & & 150 & Obtaining a user and password is considered a critical vulnerability. \\

        \arrayrulecolor{black!30}\cmidrule{1-4}
         Goal reached & & 1000 & Yield the maximum number of vulnerabilities in the environment. \\
         \bottomrule
    \end{tabular}
    \vspace{4pt}
    \caption{Reward breakdown of different information and vulnerabilities.}
    \label{tab:positive_reward}
\end{table}

\begin{table}[!htb]

    \centering
    \scriptsize
    
    \begin{tabular}{llc p{0.55\linewidth}}
        Cost Title & Configuration & Cost & Description \\
        \midrule
        \multirow{2}{*}{Crawler} & Depth search & \{1,3,4,5\} & Cost regarding the depth search for recurrent crawling in URLs. \\
        \arrayrulecolor{black!30}\cmidrule{2-4}
        & Dictionary   & \{1,2,3,4,5,6,9\} &  Cost associated to the dictionary size used for URL fuzzing and brute force\\
        \arrayrulecolor{black!30}\cmidrule{1-4}
        Parameter detection &  & \{1\} & There is only one configuration and the cost of searching for form in a given URL is low. \\
        
        \arrayrulecolor{black!30}\cmidrule{1-4}
        \multirow{3}{*}{SQL injection} & Level & \{1,2,3,4,5\} & Levels of vulnerabilities based on easy to find. \\
        \arrayrulecolor{black!30}\cmidrule{2-4}
        & Risk & \{1,2,3\} & The risk associated with this action is reflected in the cost.\\
        \arrayrulecolor{black!30}\cmidrule{2-4}
        & Techniques & \{1,3,4,2,1,1\} & The time and computational cost of the different techniques.\\
         
        \arrayrulecolor{black!30}\cmidrule{1-4}
        \multirow{2}{*}{Brute Force technique} & User dictionary & \{1,3,4,5\} & The computational cost of running a user dictionary of different sizes. \\
        \arrayrulecolor{black!30}\cmidrule{2-3}
        & Password dictionary & \{1,3,5,6,8,9\} & The cost of searching for a password dictionary of different sizes\\

        \arrayrulecolor{black!30}\cmidrule{1-4}
        Cross-site Scripting & Levels & \{2,4,6\} & The cost of running different levels of XSS scripts. \\

        \bottomrule
    \end{tabular}
    \vspace{4pt}
    \caption{Fixed Cost breakdown of different actions.}
    \label{tab:cost}
\end{table}

\end{document}